\documentclass{article}
\usepackage{amssymb}
\usepackage{amsmath}
\usepackage{amsthm}
\usepackage{amsbsy}
\usepackage{psfrag}
\usepackage{graphics}
\usepackage{graphicx}

\theoremstyle{definition}

\theoremstyle{remark}

\numberwithin{equation}{section}

\newcommand{\x}{\textbf{x}}
\newcommand{\y}{\textbf{y}}

\newcommand{\n}{\textbf{n}}
\usepackage{graphicx}
\usepackage{psfrag}
\usepackage{caption}
\usepackage{float}
\usepackage{sidecap}

\title{Tunable Double Negative Band Structure from Non-Magnetic Coated Rods}

\author{Yue Chen$^{\mbox{\tiny 1}}$
\and Robert Lipton $^{\mbox{\tiny 2}}$}
\date{}

\begin{document}
\maketitle

\baselineskip=0.9\normalbaselineskip
\vspace{-3pt}

\begin{center}
{\footnotesize 
$^{\mbox{\tiny\rm 1}}$ Department of Mathematics 
Louisiana State University
Baton Rouge, LA\\ 70803, USA. email: chenyue\symbol{'100}math.lsu.edu\\[3pt]
$^{\mbox{\tiny\rm 2}}$Department of Mathematics, Louisiana State University,
Baton Rouge, LA\\ 70803, USA. email: lipton\symbol{'100}math.lsu.edu\\[3pt]}
\end{center}

\maketitle

\begin{abstract}
A system of periodic poly-disperse coated nano-rods is considered. Both the coated nano-rods and host material are non-magnetic. The exterior nano-coating has a frequency dependent dielectric constant and the rod has a high dielectric constant.
A negative effective magnetic permeability is generated near the Mie resonances of the rods while the coating generates a negative permittivity through a field resonance controlled by the plasma frequency of the coating and the geometry of the crystal. The explicit band structure for the system is calculated in the sub-wavelength limit. Tunable pass bands exhibiting negative group velocity are generated and correspond to simultaneously negative effective dielectric permittivity and magnetic permeability. These can be explicitly controlled by adjusting the distance between rods, the coating thickness, and rod diameters.

\end{abstract}

\begin{flushleft}
Short title: Tunable Double Negative Band Structure\\
PACS 42,78
\end{flushleft}

\section{Introduction}
\label{Introduction}

Metamaterials are new class of man made materials that impart unconventional electromagnetic properties derived from sub-wavelength configurations of different conventional materials \cite{PendryHolden}. The first such materials were seen to exhibit behavior associated with negative bulk dielectric constant \cite{Pendry1998} and were constructed from a cubic lattice of metal wires. Subsequently negative effective  magnetic permeability
at microwave frequencies \cite{PendryHolden} were derived from periodic arrays of non-magnetic metallic split ring resonators \cite{PendryHolden}. Double negative or left handed metamaterials with simultaneous negative bulk permeability and permittivity at microwave frequencies have been  developed using periodic arrays of metallic posts and split ring resonators \cite{Smith}.  Subsequent work has delivered several new designs using different configurations of metallic resonators for double negative behavior \cite{17,18,19,20,21,22,23}.

For higher frequencies in the infrared and optical range an alternate strategy for constructing negative effective permeability from non-magnetic components relies on Mie resonances associated with small rods or particles made from high permittivity materials \cite{ObrienPendry}. A double negative metamaterial may be achieved by coating the high permittivity material with  a frequency dependent plasmonic or Drude type coating \cite{11} as well as  coatings with more general frequency dependence \cite{Yannopappas}. 

In this article we focus on the second approach and propose periodic assemblages of aligned non-magnetic dielectric rods each clad with a non-magnetic plasmonic coating. For this case we are able to explicitly calculate the propagation band structure in the sub-wavelength limit. The pass bands and stop bands are given by formulas that depend explicitly on the rod diameters and coating thickness.  We show how to construct tunable pass bands associated with double negative effective properties. The physical origin of the negative effective permeability is due to the excitations of Mie resonances inside the rods. On the other hand the negative effective permittivity is caused by the extreme dielectric properties near the plasma resonance of the coating. We provide the explicit relationship linking  Mie resonances and the frequency dependent effective permittivity to 
the spacing of the rods, the rod radii, and the coating thickness for the coated rod assemblage proposed here. 

These relationships are found not by appealing to effective medium theory based on the Clausius-Mossotti formula, but instead  we show that it is profitable to take a different approach and characterize wave propagation  using explicit multiscale expansions for waves inside the photonic crystal. The wave number associated with a Bloch wave inside the $d$-periodic crystal is denoted by $k=2\pi/\lambda$ where $\lambda$ is the wavelength. The approach taken here provides an explicit power series  expansion for the fields in the parameter $\eta=d k$. We outline a  systematic framework in which the homogenized dispersion relation is recovered directly from the expansion in the subwavelength  $d<<\lambda$ limit. We introduce photonic crystals made from coated rod assemblages and obtain 
explicit homogenized dispersion relations for these geometries. 
The dispersion relations provide explicit conditions on the the distance between neighboring rods, rod radii, and coating thickness  necessary for generating pass bands associated with double negative behavior. Two examples are provided showing how the band structure can be manipulated to generate double negative pass bands with negative group velocity. 

We conclude referring to related work addressing wave propagation inside high contrast media. When the crystal period and wavelength have the same length scales, the ability to open band gaps for photonic crystals is established and developed in the set of papers \cite{FK1}, \cite{FK2}, and \cite{FK3}.
The power series strategy presented here for subwavelength analysis has been utilized earlier and developed in \cite{FLS} for characterizing the  dynamic dispersion relations for Bloch waves inside plasmonic crystals and for understanding the influence of effective negative permeability on the propagation of Bloch waves inside high contrast dielectrics \cite{FLS2}. Earlier work in two scale homogenization theory for high contrast dielectrics delivers a frequency dependent effective magnetic permiability \cite{felbacqbouchette}, \cite{bouchettefelbacq} identical to the generic effective permeability tensor recovered here, see also the work of \cite{zhikov} for homogenization in high contrast media. The connection between high contrast interfaces, homogenization  and the associated generation of negative effective magnetic permeability is made in \cite{KohnShipman}. More recently 
two-scale homoginization theory has been developed for three dimensional split ring structures that deliver negative effective magnetic permeability \cite{bouchetteschwizer} and for metal fibers delivering negative effective dielectric constant \cite{bouchettebourel}. A method for creating metamaterials with prescribed effective dielectric permittivity and effective magnetic permeability at a fixed frequency
is developed in \cite{Milton2}.

\section{Electromagnetic fields inside photonic crystals made from coated rods and the subwavelength limit}
\label{derivation}
\par
\begin{figure}[h!]
\begin{center}
\includegraphics[width=0.4\textwidth]{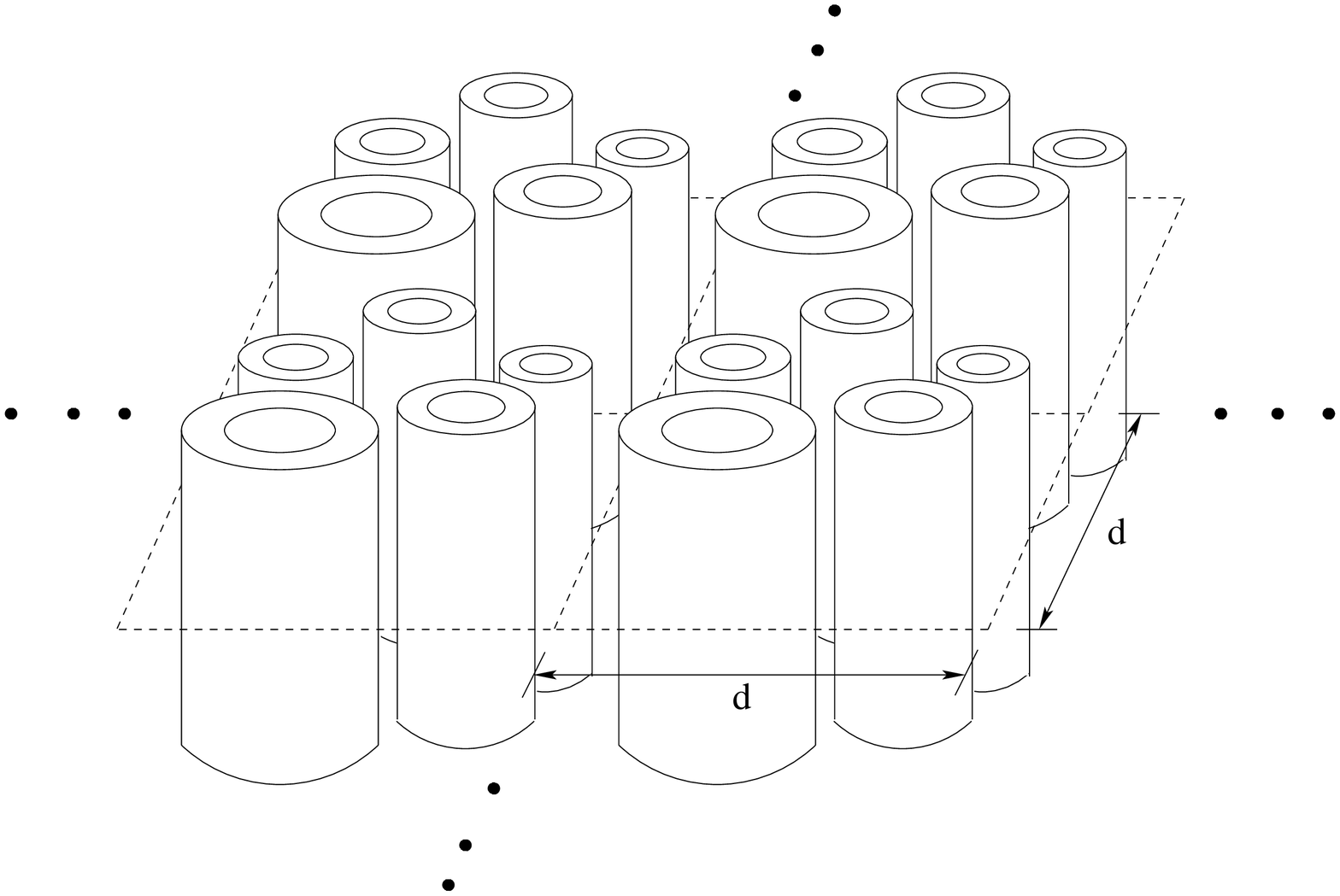}
\caption{Photonic crystal made from coated rods}
\label{rod}
\end{center}
\end{figure}

The meta material is a two-dimensional photonic crystal made of parallel coated rods, see Figure~\ref{rod}. There can be one or more coated rods inside the crystal period. The time harmonic field is p-polarized and the magnetic field inside the crystal is $\mathbf{H}=H(\x)\exp{(-i\omega t)}\mathbf{e}_3$ where $\x = (x_1, x_2)$ in the $x_1x_2$-plane. 

The period for the crystal is $d$  and the dielectric coefficient $\epsilon_d(\x)$ takes the values
\begin{equation}
\epsilon_d(\x)=
\begin{cases}
\epsilon_R & \text{ in the rod},\\
\epsilon_p &\text{ in the coating},\\
\epsilon_h &\text{ in the host material}.
\end{cases}
\end{equation}
The coating is a cylindrical shell of plasmonic material with dielectric constant $\epsilon_p(\omega^2/c^2)=1-\frac{\omega_p^2/c^2}{\omega^2/c^2}$. Here $\omega_p$ is the plasma frequency associated with the coating material and $c$ is the speed of light in vacuum. The dielectric constant of the rod is chosen according to $\epsilon_R=\frac{\epsilon_r}{d^2}$ where $\epsilon_r$ has units of area. The idea is to choose the dielectric permittivity of the rod to be large so that the corresponding Mie resonances are excited in the sub-wavelength limit. The dielectric constant of the host material is given by $\epsilon_h=1$. 

The time harmonic magnetic field for the $d$-periodic crystal is a Bloch wave 
\begin{eqnarray}
H=h(\x)\exp{(ik\hat{\kappa}\cdot\x)},
\label{bloch}
\end{eqnarray}
where $h$ is $d$-periodic in $\x$, the wave number is given by $k$, and the direction of propagation in the $x_1x_2$-plane is described by the unit vector $\hat{\kappa}=(\hat{\kappa}_1,\hat{\kappa}_2)$. $H$ satisfies the Helmholtz equation
\begin{eqnarray}
 \nabla \cdot \epsilon_d^{-1}\nabla H=-\frac{\omega^2}{c^2}H,
\label{Helmholtz}
\end{eqnarray}
with propagation bands described by dispersion relations  involving $\omega$ and $k\hat{\kappa}$. 
We  examine the situation in the sub-wavelength limit when $d$ tends to 0. Instead of developing an effective medium theory based on the Clausius-Mossotti formula  we show that it is profitable to take a different approach and expand the Bloch wave in a multiscale power series in $\eta=kd$. Here we develop a multiscale power series expansion for $h$ to recover ``homogenized sub-wavelength'' dispersion relations for plane waves inside a magnetically active double negative effective medium. 

\begin{figure}[h!]
\begin{center}
\psfrag{y1}{$y_1$}
\psfrag{y2}{$y_2$}
\psfrag{n1}{$\n$}
\psfrag{n2}{$\n$}
\psfrag{H}{$H$}
\psfrag{P}{$P$}
\psfrag{R}{$R$}
\includegraphics[width=0.4\textwidth]{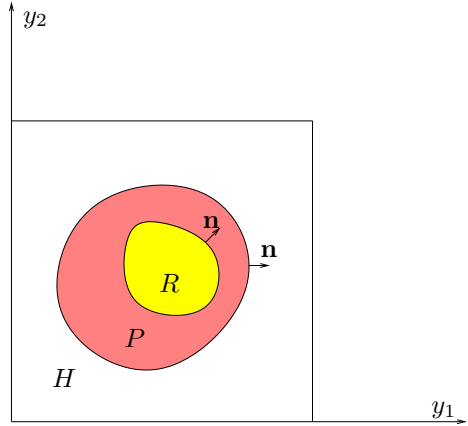}
\caption{Unit cell containing a single rod cross section. $R$ represents the rod cross section , $P$ the plasmonic coating and $H$ the host.}
\label{unitcell}
\end{center}
\end{figure}

We introduce the variable $\y=\x/d$ that takes the period cell of size $d$  into the unit period cell $Y$. Here $Y$ is split into the subdomains given by the union of rod cross sections $R$, the rod coatings $P$ and the host $H$. Figure ~\ref{unitcell}
illustrates a unit cell $Y$ containing one coated rod cross section.
The expansion parameter is given by $\eta=kd=\frac{2\pi d}{\lambda}$. The sub-wavelength limit is described by $\eta\rightarrow 0$ for $\lambda$ fixed. In what follows it is useful to introduce  $\xi$ defined by $\xi^2=\frac{\omega^2}{c^2k^2}$. The propagation equations satisfied by $h$ for $y$ in the unit period $Y$ are
\begin{eqnarray}
-Lh&=&\eta^2\xi^2 h,\hbox{ in the host},\nonumber\\
-Lh&=&\eta^2\xi^2\epsilon_P(\xi^2k^2) h, \hbox{ in the coating},\nonumber\\
-Lh&=&\epsilon_r\xi^2k^2 h, \hbox{ in the rod},\label{propagation}
\end{eqnarray}
where
\begin{eqnarray}
Lh&=&\Delta h+2i\eta\hat{\kappa}\cdot\nabla h-\eta^2 h,
\label{operatorL}\\
\epsilon_P(\xi^2k^2)&=&1-\frac{\omega_P^2/c^2}{\xi^2k^2},
\label{epsprenormalized}
\end{eqnarray}
and the transmission conditions
\begin{eqnarray}
n\cdot\left(\nabla h +i\eta\hat{\kappa} h\right)_{\scriptscriptstyle_{|_H}}&=&n\cdot\epsilon_P^{-1}(\xi^2k^2)\left(\nabla h +i\eta\hat{\kappa} h\right)_{\scriptscriptstyle_{|_C}},\hbox{ H-C interface},
\label{hostcoat}\\
n\cdot\epsilon_P^{-1}(\xi^2k^2)\left(\nabla h +i\eta\hat{\kappa} h\right)_{\scriptscriptstyle_{|_C}}&=&n\cdot\left(\frac{\eta^2}{k^2\epsilon_r}\right)\left(\nabla h +i\eta\hat{\kappa} h\right)_{\scriptscriptstyle_{|_R}}, \hbox{ R-C interface}.
\label{coatrod}
\end{eqnarray}
Here the interface separating the host phase and coating phase is denoted by H-C and the interface between the coating phase and the rod is denoted by R-C. 
Evaluation of quantities on the host side of the interface is denoted by the subscript $H$,  on the coating side by $C$, and by $R$ on the rod side. The unit normal vector $n$ on the host -- coating interface points from the coating into the host and the unit normal vector on the rod -- coating interface points from the rod into the coating. 
We expand $h$ and $\xi^2$ as
\begin{eqnarray}
 h(d\,\y)&=&h_0(\y)+\eta h_1(\y)+\eta^2 h_2(\y)+\cdots,
\label{2scaleforh1st}\\
\xi^2&=&\xi_0^2+\eta\xi_1^2+\eta^2\xi_2^2+\cdots
\label{2scaleforxi1st}
\end{eqnarray}
where each term in the expansion is continuous in $\y$ and periodic on $Y$. Substitution of the expansion into \eqref{propagation}, \eqref{hostcoat}, and \eqref{coatrod} equating like powers of $\eta$ gives the following equations used to determine $h_0$:
\begin{eqnarray}
-\Delta h_0&=&0,\hbox{ in the host},\label{eta01}\\
-\Delta h_0&=&0, \hbox{ in the coating},\label{eta02}\\
-\Delta h_0&=&\epsilon_r\xi_0^2k^2 h, \hbox{ in the rod},\label{eta03}
\end{eqnarray}
and transmission conditions,
\begin{eqnarray}
n\cdot\left(\nabla h_0\right)_{\scriptscriptstyle_{|_H}}&=&n\cdot\epsilon_P^{-1}(\xi_0^2k^2)\left(\nabla h_0\right)_{\scriptscriptstyle_{|_C}},\hbox{ H-C interface},
\label{HC0}\\
n\cdot\epsilon_P^{-1}(\xi_0^2k^2)\left(\nabla h_0\right)_{\scriptscriptstyle_{|_C}}&=&0, \hbox{ R-C interface}.
\label{CR0}
\end{eqnarray}
Noting further that $h_0$ is continuous across these interfaces we apply \eqref{eta01}, \eqref{eta02}, \eqref{HC0}, and \eqref{CR0} to discover that $h_0$ is a constant function $h_0=\overline{h}$ outside the rods. From linearity it follows that the $h_0$ field inside the rod can be written as $h_0=\overline{h}m(\y)$ where
\begin{eqnarray}
-\Delta m&=&\epsilon_r\xi_0^2k^2 m, \hbox{ in the rod},\label{eta03m}
\end{eqnarray}
and $m=1$ on the boundary of the rod. Define
the function $g(\y)$ to be $1$ for $\y$ outside the rod and $g(\y)$ to be $m(\y)$ for $\y$ inside the rod then $h_0(\y)$ is defined up to a multiplicative constant $\overline{h}$ and is given by
\begin{eqnarray}
h_0(\y)=\overline{h} g(\y).
\label{h0}
\end{eqnarray}
\par
Next we determine $h_1$ exterior to the rods. On equating like powers of $\eta$ we find that
\begin{eqnarray}
-\Delta h_1&=&0,
\label{deltah1}
\end{eqnarray}
outside the rods, and the corresponding transmission conditions for $h_1$  are given by
\begin{eqnarray}
n\cdot\left(\nabla h_1+ i\hat{\kappa} \overline{h}\right)_{\scriptscriptstyle_{|_H}}&=&n\cdot\epsilon_P^{-1}(\xi_0^2k^2)\left(\nabla h_1 +i\hat{\kappa} \overline{h}\right)_{\scriptscriptstyle_{|_C}},\hbox{ H-C interface},
\label{HC1}\\
n\cdot\epsilon_P^{-1}(\xi_0^2k^2)\left(\nabla h_1 +i\hat{\kappa} \overline{h}\right)_{\scriptscriptstyle_{|_C}}&=&0, \hbox{ R-C interface}.
\label{CR1}
\end{eqnarray}
The equation \eqref{deltah1}, transmission condition \eqref{HC1} and boundary condition \eqref{CR1} determine $h_1$ up to an additive constant $c_1$
and from the linearity of the problem we write $h_1=i\overline{h}W\cdot\hat{\kappa}+c_1$. Here  $W(\y)=(w_1(\y),w_2(\y))$ where $w_i$, $i=1,2$ are the periodic solutions of the cell problem
\begin{eqnarray}
-\Delta w_i&=&0, \hbox{for $\y$ outside the rod.}\label{eqcell1}\\
n\cdot\left(\nabla w_i+ e^i \right)_{\scriptscriptstyle_{|_H}}&=&n\cdot\epsilon_P^{-1}(\xi_0^2k^2)\left(\nabla w_i +e^i\right)_{\scriptscriptstyle_{|_C}},\hbox{ H-C interface},
\label{HC1i}\\
n\cdot\epsilon_P^{-1}(\xi_0^2k^2)\left(\nabla w_i +e^i \right)_{\scriptscriptstyle_{|_C}}&=&0, \hbox{ R-C interface},
\label{CR1i}
\end{eqnarray}
where $e^1=(1,0)$ and $e^2=(0,1)$.
We  define the region inside the period cell exterior to the rods by $D$ and introduce the position dependent dielectric constant $\epsilon(\y)$ taking the value $1$ inside the host and $\epsilon_P(k^2\xi_0^2)$ inside the coating.
For future reference we introduce the tensor $\epsilon_{eff}^{-1}(\xi_0^2k^2)$ defined by
\begin{eqnarray}
\epsilon_{eff}^{-1}(\xi_0^2k^2)=\int_D\epsilon(\y)\left(\nabla W+I\right)d\y,
\label{eff}
\end{eqnarray}
where $I$ is the $2\times2$ identity tensor.

Now we identify the effective magnetic permeability. Following \cite{PendryHolden}, \cite{ObrienPendry} the average component of the  $B$ field parallel to the rods is given by
\begin{eqnarray}
B_{ave}=\frac{1}{d^2}\int_{Y_d}h(\x)e^{{i(k\hat{\kappa}\cdot\x})}dx_1dx_2.
\label{averageB}
\end{eqnarray}
Here $Y_d$ is a period cell for the $d$-periodic crystal.
Substituting the expansion \eqref{2scaleforxi1st} for $h$ into \eqref{averageB}, retaining the lowest order terms and taking $d\rightarrow 0$ gives
\begin{eqnarray}
\lim_{d\rightarrow 0}B_{ave}=\overline{h}\int_Y g(\y)dy_1dy_2.
\label{averagedB}
\end{eqnarray}
The average component of the $H$ field parallel to the rods is given by \cite{PendryHolden}, \cite{ObrienPendry}
\begin{eqnarray}
H_{ave}=\frac{1}{d}\int_{(0,0,0)}^{(0,0,d)}h(\x)e^{{i(k\hat{\kappa}\cdot\x})}dx_1dx_2,
\label{averageH}
\end{eqnarray}
here the average is taken over any line parallel and exterior to the rods. Sending $d\rightarrow 0$ gives
\begin{eqnarray}
\lim_{d\rightarrow 0}H_{ave}=\overline{h}.
\label{averagedH}
\end{eqnarray}
Using \eqref{averagedB} and \eqref{averagedH} the effective permittivity    is given by
\begin{eqnarray}
\lim_{d\rightarrow 0}\frac{B_{ave}}{H_{ave}}=\mu_{eff}(\xi_0^2k^2).
\label{effectivemag}
\end{eqnarray}
We recover the explicit form of the effective permeability $\mu_{eff}(\xi_0^2k^2)$ for a distribution of circular rods inside the unit period cell. 
To do so recall the following Dirichlet eigenvalues $E$ and eigenfunctions $\psi$ for each rod cross section given by 
\begin{eqnarray}
-\Delta \psi=E\psi  \text{  for  } \y \text{ inside rod }  \quad \text{  and   } \quad \psi=0 \text{  for  } \y \text{ on rod boundary }.
\label{eigendirich}
\end{eqnarray}
For rods with circular cross section of radius $0<R_r<1$ and polar coordinates $0<\rho<R_r$, $0\leq\theta<2\pi$,
let $\chi_{nm}$ be the $m$th zero of the $n$th Bessel function, then the Dirichlet eigenvalues  are $E_{nm}=(\chi_{nm}/R_r)^2$, and the corresponding eigenfunctions are $\{J_n(\frac{\chi_{nm}}{R_r}\rho)e^{in\theta}, J_n(\frac{\chi_{nm}}{R_r}\rho)e^{-in\theta}\}$. The associated normalized eigenfunctions are denoted by $\psi_{nm}^{\pm}$. Notice that the eigenvalue $E_{0m}$ has only one associated normalized eigenfunction $\psi_{0m}=\frac{J_0(\chi_{0m}(\rho/R_r))}{\sqrt{\pi}R|J'_0(\chi_{0m})|}$ and $\langle\psi_{0m}\rangle^2=\frac{4\pi R^2}{\chi_{0m}^2}$ when $n=0$.
Here $\langle\cdot\rangle$ denotes the area integral over a rod cross section. It is obvious that $\langle\psi_{nm}^{\pm}\rangle^2=0$ for $n\neq 0$.
\par
Expanding $m(\y)$ in the basis $\psi_{nm}^{\pm}$ noting that $m-1=0$ outside the rods gives for each rod
\begin{eqnarray}
m=1-\sum_{m=0}^\infty\frac{\xi_0^2k^2}{\xi_0^2k^2-\frac{E_{0m}}{\epsilon_r}}\langle\psi_{0m}\rangle\psi_{0m}.
\label{solutionofm}
\end{eqnarray}
For a distribution of rods with radii ${R_r}_j$, $j=1,\ldots,\ell$, we have 
\begin{eqnarray}
\mu_{eff}(\xi_0^2k^2)=\int_Y g(\y)d\y=1-\sum_{j=0}^\ell N(R_{rj})\sum_{m=0}^\infty\frac{4\pi R_{rj}^2}{\chi_{0m}^2}\frac{\xi_0^2k^2}{\xi_0^2k^2-k_{jm}^2} ,
\label{explicitexpression}
\end{eqnarray}
where $N(R_{rj})$ is the number of rods with radius $R_{rj}$ and $k_{jm}^2=\frac{\chi_{0m}^2}{\epsilon_r R_{rj}^2}$. 

Last we find the homogenized sub-wavelength dispersion relation. Recall that $\xi^2k^2=\frac{\omega^2}{c^2}$ and from \eqref {2scaleforxi1st}
\begin{eqnarray}
\frac{\omega^2}{c^2}&=&\frac{\omega_0^2}{c^2}+\eta\frac{\omega_1^2}{c^2}+\eta^2\frac{\omega_2^2}{c^2}+\cdots,
\label{2scaleforomega1st}
\end{eqnarray}
where
\begin{eqnarray}
\frac{\omega_i^2}{c^2}&=&\xi_i^2k^2,\hbox{ for $i=0,1,\ldots$}.
\label{dispersionprep}
\end{eqnarray}
The sub-wavelength dispersion relation identified as the dispersion relation between $\omega_0$ and $k\hat{\kappa}$ given by $\frac{\omega_0^2}{c^2}=\xi_0^2k^2$. We now recover this dispersion relation by deriving the formula for $\xi_0^2$. Equating like powers of $\eta$ gives the following equations for $h_2$ exterior to the rods.
\begin{eqnarray}
-\left(\Delta h_2+2i\eta\hat{\kappa}\cdot\nabla h_1-\eta^2 \overline{h}\right)&=&\xi_0^2 \overline{h},\hbox{ in the host},\label{propagationh2a}\\
-\left(\Delta h_2+2i\eta\hat{\kappa}\cdot\nabla h_1-\eta^2 \overline{h}\right)&=&\xi_0^2\epsilon_P(\xi_0^2k^2) \overline{h}, \hbox{ in the coating},\label{propagationh2b}
\end{eqnarray}
and the transmission conditions
\begin{eqnarray}
&&\left(\xi_0^2k^2-\frac{\omega_p^2}{c^2}\right)n\cdot\left(\nabla h_2 +i\hat{\kappa} h_1\right)_{\scriptscriptstyle_{|_H}}+\left(\xi_1^2k^2\right)n\cdot\left(\nabla h_1 +i\hat{\kappa} \overline{h}\right)_{\scriptscriptstyle_{|_H}}\nonumber\\
&&=\left(\xi_0^2k^2\right)n\cdot\left(\nabla h_2 +i\hat{\kappa} h_1\right)_{\scriptscriptstyle_{|_C}}
+\left(\xi_1^2k^2\right)n\cdot\left(\nabla h_1 +i\hat{\kappa} \overline{h}\right)_{\scriptscriptstyle_{|_C}}
,\hbox{ H-C interface},
\label{hostcoat2}\\
&&\left(\xi_0^2k^2\right)n\cdot\left(\nabla h_2 +i\hat{\kappa} h_1\right)_{\scriptscriptstyle_{|_C}}+\left(\xi_1^2k^2\right)n\cdot\left(\nabla h_1 +i\hat{\kappa} \overline{h}\right)_{\scriptscriptstyle_{|_C}}+\left(\xi_2^2k^2\right)n\cdot\nabla {h_0}_{\scriptscriptstyle_{|_C}}\nonumber\\
&&=\frac{1}{k^2\epsilon_r}\left(\xi_0^2k^2-\frac{\omega_p^2}{c^2}\right)n\cdot\nabla {h_0}_{\scriptscriptstyle_{|_R}}, \hbox{ R-C interface}.
\label{coatrod2}
\end{eqnarray}
Integrating and adding \eqref{propagationh2a} and \eqref{propagationh2b} gives the solvability condition
\begin{eqnarray}
\xi_0^2\int_{D} \overline{h}\,d\y&=&-\int_H\left(\Delta h_2+2i\hat{\kappa}\cdot\nabla h_1-\overline{h}\right)\,d\y\nonumber\\
&-&\int_P\epsilon^{-1}(\xi_0^2k^2)\left(\Delta h_2+2i\hat{\kappa}\cdot\nabla h_1-\overline{h}\right)\,d\y.
\label{solve}
\end{eqnarray}
On integrating by parts and applying \eqref{eta03} together with the transmission conditions\eqref{CR0},  \eqref{HC1}, \eqref{CR1},  \eqref{hostcoat2}, and \eqref{coatrod2} shows that  \eqref{solve} is equivalent to
\begin{eqnarray}
\xi_0^2=\mu_{eff}^{-1}(\xi_0^2k^2)\epsilon_{eff}^{-1}(\xi_0^2k^2)\hat{\kappa}\cdot\hat{\kappa}.
\label{dispersionhomog}
\end{eqnarray}
Writing $\omega_0^2/c^2=\xi_0^2k^2$ and substitution into \eqref{dispersionhomog} delivers the homogenized subwavelength dispersion relation
\begin{eqnarray}
\frac{\omega_0^2}{c^2}=n_{eff}^{-2}k^2,
\label{subwavedis}
\end{eqnarray}
where the effective index of diffraction $n_{eff}^2$ is given by
\begin{eqnarray}
n_{eff}^2=\mu_{eff}(\frac{\omega_0^2}{c^2})\left(\epsilon_{eff}^{-1}(\frac{\omega_0^2}{c^2})\hat{\kappa}\cdot\hat{\kappa}\right)^{-1}.
\label{index}
\end{eqnarray}

Applying \eqref{2scaleforh1st} and collecting results shows that to lowest order the 
Bloch waves $H=h(\x)e^{i(k\hat{\kappa}\cdot\x-t\omega)}$ inside the photonic crystal are given by the subwavelength expansion
\begin{eqnarray}
H(\x,t)=\left(\overline{h}g(\frac{\x}{d})+\eta h_1(\frac{\x}{d})+O(\eta^2)\right)\exp{i(k\hat{\kappa}\cdot\x-t(\omega_0+\eta\omega_1+O(\eta^2))}.
\label{blochexpanded}
\end{eqnarray}

In the subwavelength limit the spatial averages of $H$ correspond to spatial averages of plane waves associated with a magnetically active effective medium. To see this take any planar averaging domain $S$ and pass to the $d\rightarrow 0$ limit in the average to get
\begin{eqnarray}
\lim_{d\rightarrow 0}\frac{1}{Area(S)}\int_S H(\x,t)dx_1 d x_2&=&\lim_{d\rightarrow 0}\frac{1}{Area(S)}\int_S \overline{h}g(\frac{\x}{d})\exp{i(k\hat{\kappa}\cdot\x-t\omega_0)}dx_1 d x_2\nonumber\\
&=&\frac{1}{Area(S)}\int_S \mu_{eff}\overline{h}\exp{i(k\hat{\kappa}\cdot\x-t\omega_0)}dx_1 dx_2.
\label{limareaavg}
\end{eqnarray}
In a similar way, taking averages over any interval $a<x_3<b$ on any line parallel to the $x_3$ axis not intersecting the coated rods gives
\begin{eqnarray}
\lim_{d\rightarrow 0}\frac{1}{b-a}\int_{(0,0,a)}^{(0,0,b)} H(\x,t)dx_3&=&\lim_{d\rightarrow 0}\frac{1}{b-a}\int_{(0,0,a)}^{(0,0,b)} \overline{h}g(\frac{\x}{d})\exp{i(k\hat{\kappa}\cdot\x-t\omega_0)}dx_3\nonumber\\
&=&\frac{1}{b-a}\int_{(0,0,a)}^{(0,0,b)} \overline{h}\exp{i(k\hat{\kappa}\cdot\x-t\omega_0)}dx_3.
\label{limarelavg}
\end{eqnarray}
Thus the appropriate averages of the plane waves $H_{hom}=\overline{h}\exp{i(k\hat{\kappa}\cdot\x-t\omega_0)}$ and\\ $B_{hom}=\mu_{eff}\overline{h}\exp{i(k\hat{\kappa}\cdot\x-t\omega_0)}$ provide approximations to the average field seen in the $d$ periodic photonic crystal for $0<d<<k$.

\section{Electromagnetic fields inside coated rod assemblages}
\label{assemblages}

\begin{figure}[h!]
\begin{center}
\begin{psfrags} 
\psfrag{R1}{$R_r$}
\psfrag{R2}{$R_p$}
\psfrag{R3}{$R_h$}
\includegraphics[width=0.2\textwidth]{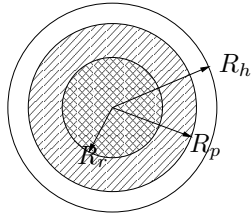}
\end{psfrags} 
\caption{Doubly Coated Cylinder}
\label{doublycoatedcylinder}
\end{center}
\end{figure}

Here we introduce a special class of photonic crystals made from coated rod assemblages and derive an explicit formula for the subwavelength dispersion relation \eqref{subwavedis}. The formula shows how the band structure depends explicitly on the distribution of rod radii, coating thickness and distance between neighboring rods. The assemblage is described as follows. Consider first the unit period cell $Y$ filled with disks with radii ranging down to the infinitesimal. Inside each disk we place a centered rod cross section then a concentric coating of plasmonic material and finally a concentric coating of host material, see Figure ~\ref{doublycoatedcylinder}.
\par
\begin{figure}[h!]
\begin{center}
\begin{psfrags} 
\includegraphics[width=0.2\textwidth]{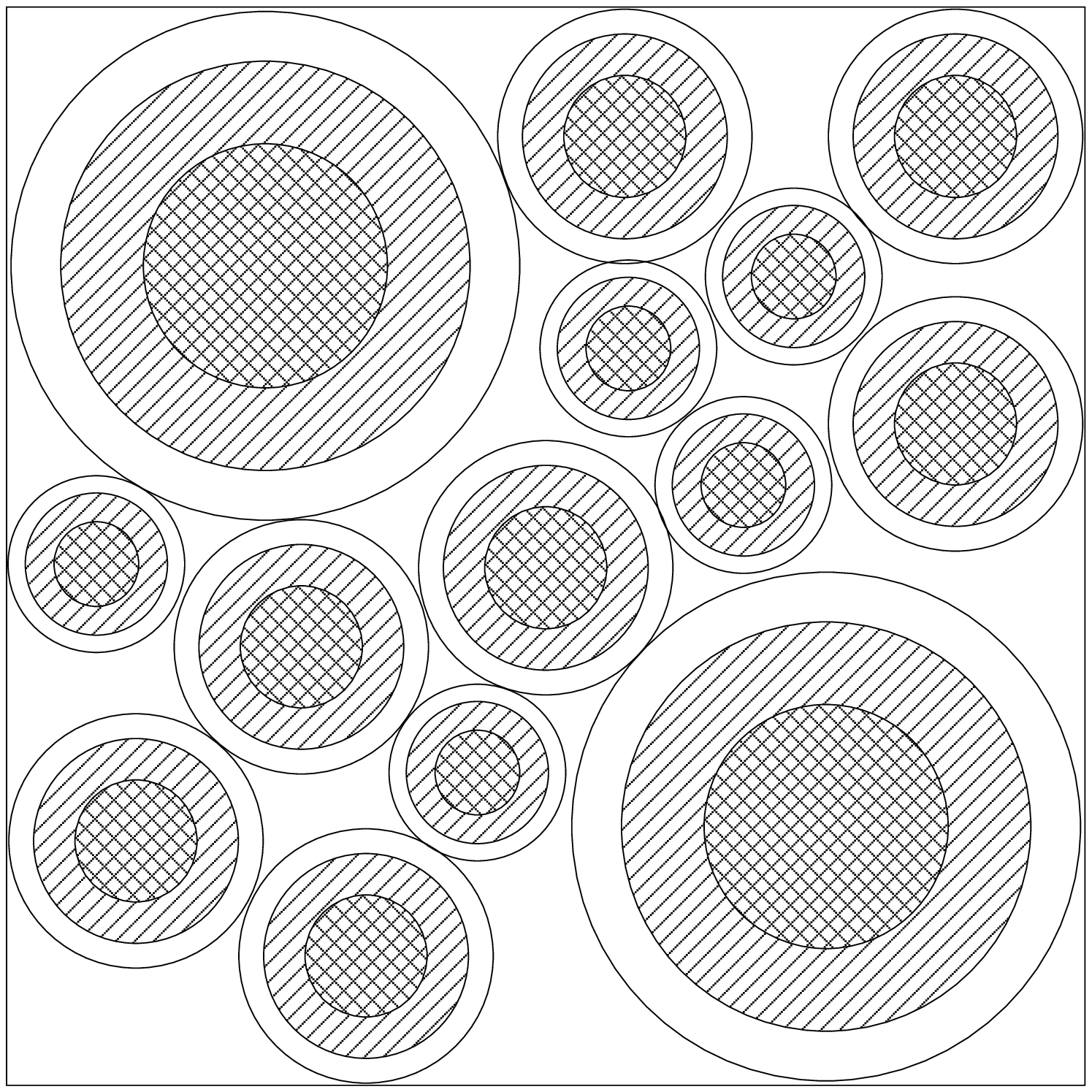}
\end{psfrags} 
\caption{Cross section of the unit period filled with coated rods. The coated rods fill all space.}
\label{crosssection}
\end{center}
\end{figure}

The cross sections of the coated rods together with the connected host material is depicted in Figure \ref{crosssection}. The ratio of the radii of the rod cross section, plasmonic coating, and host coating are the same for all the disks and we denote the area fractions of the host, coating, and core phases by $\theta_H$, $\theta_P$, and $\theta_R$ respectively, see Figure \ref{doublycoatedcylinder}. The radii of the rods used in the assemblage is 
denoted  by  $R_{r1}>R_{r2}>R_{r3}\cdots$. The number of rods having radius $R_{rj}$ are denoted by $N(R_{rj})$. For rods of radii $R_{rj}$ the outer radii of the core is denoted by $R_{pj}$ and the distance to the nearest neighboring rod is $R_{hj}$. for the coated rod assemblage it is clear that $R_{hj}$ is determined by $R_{rj}$ together with $\theta_R$ and  $\theta_P$.
\par
This type of configuration is known in the composites literature and is referred to as a doubly coated cylinder assemblage \cite{Schulgasser}, \cite{Milton}, \cite{LurieCherkaev}, \cite{Milgrom}. 
\begin{figure}[h!]
\begin{center}

\begin{psfrags} 
\psfrag{mu}{$\mu_{eff}$}
\psfrag{wc}{$(\frac{\omega_0}{c})^2$}
\psfrag{k1}{$k^2_{00}$}
\psfrag{k2}{$k^2_{10}$}
\psfrag{k3}{$k^2_{01}$}
\psfrag{c}{$\cdots$}
\includegraphics[width=0.3\textwidth]{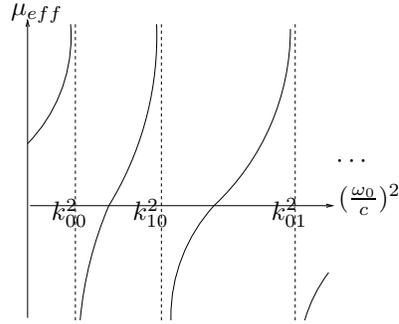}
\end{psfrags} 
\caption{the relation between $\mu_{eff}$ and $(\omega_0/c)^2$}
\label{mueffectivewc}
\end{center}
\end{figure}
The explicit formula for  $\mu_{eff}$ follows from \eqref{explicitexpression} and is given by
\begin{eqnarray}
\mu_{eff}\left(\frac{\omega_0^2}{c^2}\right)=1-\sum_{j=0}^\infty N(R_{rj})\sum_{m=0}^\infty\frac{4\pi R_{rj}^2}{\chi_{0m}^2}\frac{(\omega_0/c)^2}{(\omega_0/c)^2-k_{jm}^2}.
\label{explicitexpressionearly}
\end{eqnarray}
The poles of $\mu_{eff}(\omega_0^2/c^2)$ are given by $k_{jm}^2$, see Figure ~\ref{mueffectivewc}.

The explicit formula for the effective dielectric permittivity   follows from a standard calculation using \eqref{eqcell1},  \eqref{eqcell1}, \eqref{HC1i} and  \eqref{CR1i} (see, for example \cite{MiltonBook}) and is given by
\begin{eqnarray}
\epsilon_{eff}^{-1}=\epsilon_H^{-1}+\frac{2(1-\theta_H)\epsilon_H^{-1}}{\theta_H-\frac{2\epsilon_H^{-1}}{\epsilon_H^{-1}-\epsilon_C^{-1}}}
\label{outerercoat}
\end{eqnarray}
with
\begin{eqnarray}
\epsilon_{C}^{-1}=\epsilon_P^{-1}+\frac{2\theta_R\epsilon_P^{-1}}{2\theta_R-\theta_P}.
\label{innercoat}
\end{eqnarray}
On writing $\epsilon_p(\omega_0^2/c^2)=1-\frac{\omega_p^2/c^2}{\omega_0^2/c^2}$ and $\epsilon_H=1$ we obtain the formula for the frequency dependent effective dielectric  tensor given by
\begin{eqnarray}
\epsilon_{eff}\left(\frac{\omega_0^2}{c^2}\right)=\frac{1-(\theta_R+\theta_P)}{1+\theta_R+\theta_P}\left(\frac{\frac{1+\theta_R+\theta_P}{1-(\theta_R+\theta_P)}+\left(1-\frac{(\omega_p/c)^2}{(\omega_0/c)^2}\right)^{-1}\left(\frac{\theta_P}{2\theta_R+\theta_P}\right)}{\frac{1-(\theta_R+\theta_P)}{1+\theta_R+\theta_P}+\left(1-\frac{(\omega_p/c)^2}{(\omega_0/c)^2}\right)^{-1}\left(\frac{\theta_P}{2\theta_R+\theta_P}\right)}\right).
\label{effpermittivity}
\end{eqnarray}
The dependence of $\mu_{eff}$ and $\epsilon_{eff}$ on the rod diameter and coating thickness and host area fraction is explicitly given by \eqref{explicitexpressionearly} and \eqref{effpermittivity}. This explicit dependence allows for the tuning of the dispersion relation 
\begin{eqnarray}
k^2=\frac{\omega_0^2}{c^2}n_{eff}^{2},
\label{coateddispersion}
\end{eqnarray}
where
\begin{eqnarray}
n_{eff}^{2}=\mu_{eff}\left(\frac{\omega_0^2}{c^2}\right)\epsilon_{eff}\left(\frac{\omega_0^2}{c^2}\right).
\label{effindexproduct}
\end{eqnarray}
For each coated rod with radius $R_{rj}$, outer coating radius $R_{pj}$ and nearest neighbor distance $R_{hj}$
the $h_0$ field inside the rod is given by \eqref{solutionofm} and the $h_1$ field outside the rod is
\begin{eqnarray}
&&h_1=(1+\theta_R)^{-1}\left(r\cos{\gamma}+\frac{R_{rj}^2}{r}\cos{\gamma}\right), \hbox{ $R_{rj}<r<R_{pj}$},
\label{coath1}\\
&&h_1=\frac{2\theta_R+\theta_P+\theta_P\left(1-\frac{\omega_p^2/c^2}{\omega_0^2/c^2}\right)^{-1}}{F}r \cos{\gamma}+\nonumber\\
&&+\frac{2\theta_R+\theta_P-\theta_P\left(1-\frac{\omega_p^2/c^2}{\omega_0^2/c^2}\right)^{-1}}{F}\frac{R_{pj}^2}{r} \cos{\gamma},\hbox{ $R_{pj}<r<R_{hj}$},
\label{hosth1}
\end{eqnarray}
where $(r,\gamma)$ are local polar coordinates inside the coated rod and $F$ is given by
\begin{eqnarray}
F=(2\theta_R+\theta_P)(1+\theta_R+\theta_P)+\left(1-\frac{\omega_p^2/c^2}{\omega_0^2/c^2}\right)^{-1}\theta_P(1-(\theta_R+\theta_P)).
\label{DD}
\end{eqnarray}
It is clear that $h_0$ diverges inside the rods with radii $R_{rj}$  as $\omega_0^2/c^2$ approaches $k_{jm}^2$ while $h_1$ diverges
in the neighborhoods of all the coated rods, $R_{pj}<r<R_{hj}$ as $\omega_0^2/c^2$ approaches the zero of $\epsilon_{eff}\left(\frac{\omega_0^2}{c^2}\right)$.

\section{Tunable double negative behavior}
\label{negative}
\begin{figure}[h!]
\begin{center}
\begin{psfrags} 
\psfrag{epsilon}{$\epsilon_{eff}$}
\psfrag{wc}{$(\frac{\omega_0}{c})^2$}
\psfrag{p}{$p$}
\psfrag{pstar}{$p^*$}

\includegraphics[width=0.3\textwidth]{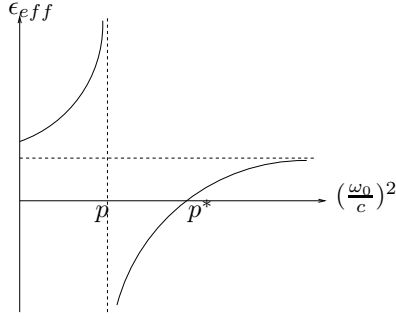}
\end{psfrags} 
\caption{the relation between $\epsilon_{eff}$ and $(\omega_0/c)^2$}
\label{epsiloneffectivewc}
\end{center}
\end{figure}

When $\epsilon_{eff}$ and $\mu_{eff}$  have the same sign it is clear that $n^{2}_{eff}>0$ and wave propagation occurrs.  In this section, we show that it is possible to tune the assemblage through the choice of rod radii, coating thickness and host volume fraction to exhibit single negative, double negative, and double positive behavior depending on the signs of  $\epsilon_{eff}$ and $\mu_{eff}$. For coated cylinder assemblages it is evident that the distance between neighboring rods and the coating thickness is explicitly determined by the area fraction of the host phase, the area fraction of the rods and the rod radii $R_{rj}$.
The dielectric function \eqref{effpermittivity} has only one  pole and one zero and these depend explicitly on the geometry of the coated rod assemblage and are given by
\begin{eqnarray}
p=(\frac{\omega_p}{c})^2\frac{\theta_H(2\theta_R+\theta_P)}{\theta_H(2\theta_R+\theta_P)+(2-\theta_H)\theta_P}
\end{eqnarray}
and 
\begin{eqnarray}
p^*=(\frac{\omega_p}{c})^2\frac{(2-\theta_H)(2\theta_R+\theta_P)}{(2-\theta_H)(2\theta_R+\theta_P)+\theta_H\theta_P} .
\end{eqnarray}
Observe that $p<p^*$ and $\epsilon_{eff}$ is negative when $p< (\omega_0/c)^2 < p^*$ . The relation between $\epsilon_{eff}$ and $(\omega_0/c)^2$ is displayed in Figure ~\ref{epsiloneffectivewc}. 
\par
From Figure ~\ref{mueffectivewc} and Figure ~\ref{epsiloneffectivewc},it is clear that both $\epsilon_{eff}$ and $\mu_{eff}$ are negative for $ p< (\omega_0/c)^2 < p^*$ provided that $({\omega_0}/{c})^2$ is simultaneously greater than but close to the poles of $\mu_{eff}$. Therefore  for $\epsilon_{eff}$ and $\mu_{eff}$ to be simultaneously negative it is required that  $p< k_{jm}^2 < p^*$, i.e., the assemblage must contain  rods with radii
$R_{rj}$ satisfying the condition
\begin{eqnarray}
\frac{\chi_{0m}^2c^2}{\epsilon_r\omega_p^2}\frac{(2-\theta_h)(2\theta_r+\theta_p)+\theta_h\theta_p}{(2-\theta_h)(2\theta_r+\theta_p)}<R_{rj}^2<\frac{\chi_{0m}^2c^2}{\epsilon_r\omega_p^2}\frac{\theta_h(2\theta_r+\theta_p)+(2-\theta_h)\theta_p}{\theta_h(2\theta_r+\theta_p)}.
\label{tuningnegative}
\end{eqnarray}

\begin{figure}[h] 
\begin{center}
\begin{psfrags} 
\psfrag{w}{$(\omega_0 /c)^2 $}
\psfrag{k}{$k^2$}
\psfrag{p}{$p$}
\psfrag{k1}{${k}^2_{00}$}
\psfrag{k2}{${k}^2_{10}$}
\psfrag{k3}{${k}^2_{01}$}
\psfrag{pstar}{$p^*$}
\psfrag{BandGap}{\hbox{Stop Band}}
\psfrag{c}{$\vdots$}
\includegraphics[width=0.3\textwidth]{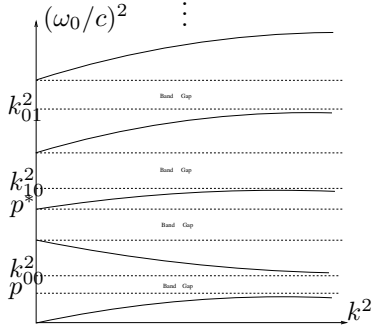}
\end{psfrags} 
\caption{Dispersion relation $k^2=(\frac{\omega_0 }{c})^2n^2_{eff}$ for a material with $\epsilon_p=1-\frac{\omega_p^2/c^2}{\omega_0^2/c^2}$ where $\omega_p=25.8$ THz. The assemblage contains rods of radii $R_{r0}=0.2$ and $R_{r1}=0.17$. }
\label{example1}
\end{center}
\end{figure}

Now we fix the area fractions $\theta_R=0.8 , \theta_P=0.1, \theta_H=0.1$ , and  set $\omega_p=25.8$ THz and $\epsilon_r=200 m^2 $ so that
$p=0.349\times 10^{10}$ $m^{-2}$ and $p^*=0.737\times 10^{10}$ $m^{-2}$.
The dispersion relation $k^2=(\frac{\omega}{c})^2n^{2}_{eff}$ is displayed in Figure ~\ref{example1} for an assemblage with the two largest rod radii chosen to be $R_{r0}=0.2$ and $R_{r1}=0.17$. These radii satisfy the requirement given by \eqref{tuningnegative} while the radii of all other rods in the assemblage are chosen to lie below these two values and do not satisfy \eqref{tuningnegative}. It is clear from the figure that, $\epsilon_{eff}<0$  when $p<({\omega_0/c})^2<p^*$ and otherwise  $\epsilon_{eff}\geq 0$ . In this example,  $p$ is below ${k}_{00}^2$ and $p^*$ lies between $k_{00}^2$ and $k_{10}^2$ . Notice in Figure ~\ref{example1} that the slope is negative for the dispersion curve lying between $k_{00}^2$ and $p^*$, indicating a negative group velocity.
\par
\begin{figure}[h!]
\begin{center}
\begin{psfrags} 
\psfrag{w}{$(\omega_0 / c)^2 $}
\psfrag{k}{$k^2$}
\psfrag{p}{$p$}
\psfrag{pstar}{$p^*$}
\psfrag{k1}{${k}^2_{00}$}
\psfrag{k2}{${k}^2_{10}$}
\psfrag{k3}{${k}^2_{01}$}
\psfrag{c}{$\vdots$}
\includegraphics[width=0.3\textwidth]{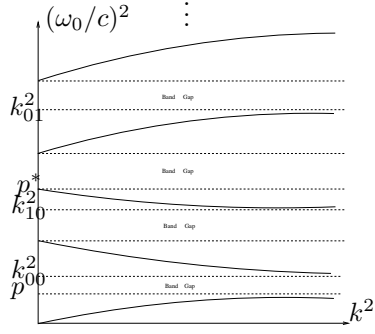}
\end{psfrags} 
\caption{Dispersion relation $k^2=(\frac{\omega_0}{c})^2n^2_{eff}$ for a material with $\epsilon_p=1-\frac{\omega_p^2/c^2}{\omega_0^2/c^2}$ where $\omega_p=25.8$ THz. The largest two rod radii appearing in the assemblage are given by $R_{r0}=0.2$ and $R_{r1}=0.18$. } 
\label{example2}
\end{center}
\end{figure}

In Figure ~\ref{example2}, we choose the largest two radii to be given by $R_{r0}=0.2$ and $R_{r1}=0.18$   and all other rods have radii below these values. For this case, $p$ is below $k_{00}^2$ but $p^*$ is between ${k}_{10}^2$ and ${k}_{01}^2$.
\par

\begin{figure}[h!]
\begin{center}

\includegraphics[width=0.4\textwidth]{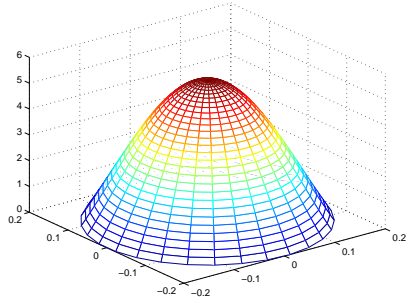}

\caption{ the graph of $\psi_{00}$} 
\label{graphofpsi00}
\end{center}
\end{figure}

\begin{figure}[h!]
\begin{center}

\includegraphics[width=0.4\textwidth]{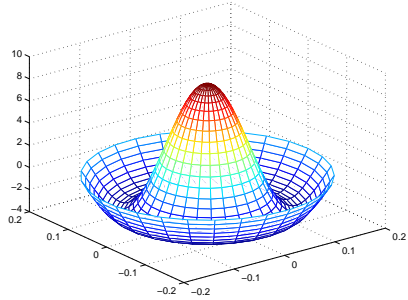}

\caption{ the graph of $\psi_{01}$} 
\label{graphofpsi01}
\end{center}
\end{figure}

Notice that for points $\x$ inside the coating phase and for $(\omega_0/c)^2$ close to $k^2_{0m}$, that \eqref{solutionofm} and  \eqref{blochexpanded} show that $H(\x,t)$ behaves like $h(\x)e^{i(k\hat{\kappa}\cdot\x-t\omega)}$  where 
\begin{eqnarray}
h(\x)=\mu_{eff}^{-1}\overline{h}e^{ik\hat{\kappa}\cdot\x}(1-\frac{(\omega_0/c)^2}{(\omega_0/c)^2-k^2_{0m}}\langle1|\psi_{0m}\rangle)\psi_{0m}.
\label{profile}
\end{eqnarray}
In particular, when $(\omega_0/c)^2$ is near $k^2_{00}$ , the profile of $h$ is  $\psi_{00}$ (see Figure ~\ref{graphofpsi00} with $R_{r0}=0.2$ ) and when $(\omega_0/c)^2$ is near $k^2_{01}$ , the profile of $h$ is $\psi_{01}$ (see Figure ~\ref{graphofpsi01} with $R_{r0}=0.2$ ).


\section{Acknowledgements}
This research is supported by NSF grant DMS-0807265 and AFOSR grant FA9550-05-0008.

\end{document}